\documentclass[12pt]{article}
\usepackage{epsf}
\usepackage{epsfig}

\usepackage{epsf}

\begin{document}

\title{Gravitational Waves from a Pulsar Kick Caused by Neutrino 
Conversions}
\author{Lee C. Loveridge \thanks{leecl@physics.ucla.edu} \\ 
           {\small \it Department of Physics and Astronomy} \\ 
	   {\small \it UCLA, Los Angeles, CA 90095-1547}}
\date {UCLA/03/TEP/25 \\ \today}
\maketitle

\begin{abstract}

It has been suggested that the observed pulsar velocities are caused by an
asymmetric neutrino emission from a hot neutron star during the first
seconds after the supernova collapse.  We calculate the magnitude of
gravitational waves produced by the asymmetries in the emission of
neutrinos.  The resulting periodic gravitational waves may be detectable 
by LIGO and LISA in the event of a nearby supernova explosion. 

\end{abstract}

\section{Introduction}
It has been suggested that the proper motions of pulsars~\cite{astro} may
be caused by a slight assymetry in the ejection of
neutrinos~\cite{OscKicks,sterile,sterile_dark}, which carry away most of
the energy of a supernova.  In fact, it would take only a 1 \% assymetry in
the neutrino emmission to achieve the observed neutron star velocities.
Such an assymetry could be caused by neutrino oscillations in matter, in a
strong magnetic field.  Although neutrino magnetic moments are negligible,
the neutrino interactions with matter depend on the polarization of matter
fermions~\cite{polar}.  The viability of the neutrino kick mechanism has
been discussed at length in several papers
\cite{OscKicks,sterile,sterile_dark}.  This mechanism can work if there is
a sterile neutrino with mass 1-20 keV and a small, $\sin \theta \sim
10^{-4}$, mixing with the electron neutrino.  It is intriguing, that the
same neutrino could make up the cosmological dark
matter~\cite{sterile_dark}.  Unfortunately, detection of such a neutrino in
a terrestrial experiment is probably not feasible at present because the
mixing angle $\sin \theta $ is very small. However, in the event of a
nearby supernova, the asymmetric emission of neutrinos from a rotating
nascent neutron star can produce gravity waves that may be detectable.

In this paper we will describe the expected gravity waves (GW) from a
pulsar kick due to neutrino conversions taking into account the rotation of
the neutron star and the neutrino asymmetry associated with this
mechanism. Detection of such a signal can help understand the 
origin of the pulsar kicks. 

Gravitational waves from the supernova neutrinos have been considered in the
literature\cite{Epstein,Burrows,Cuesta}.  However, the source geometry was
assumed somewhat different from that specific to the neutrino kick
mechanism.  Epstein~\cite{Epstein} has calculated the burst radiation
assuming the distribution of neutrinos is roughly elliptical.  The assymetry
was assumed to arise due to flattening by centripetal motion.  Burrows
and Hayes~\cite{Burrows} performed a hydrodynamic simulation of a
supernova with some imposed initial assymetry.  Cuesta~\cite{Cuesta}
described the burst radiation from the sudden neutrino conversions. 
He has also proposed that the GWs themselves may be the source of the 
neutron star momentum kick, through their radiation back
reaction~\cite{Cuesta2}.

However, there is an additional, and possibly stronger, source of
gravitational radiation due to the asymmetric emission of neutrinos from a
rapidly rotating neutron star.  One can think of the escaping neutrinos as
forming an off-centered rotating beam, at some angle to the rotation axis,
superimposed on a spherically symmetric distribution.  The symmetric
component does not
produce gravity waves, but the 1\% of all neutrinos that make up
the asymmetric ``beam'' are a source of a periodic GW signal.

\section{Gravity waves from a spinning neutrino beam}
We will adopt Epstein's notation~\cite{Epstein}, with a few minor
modifications.  We define the nontensor potentials
\begin{eqnarray}
h_{\mu \nu} & \equiv & g_{\mu \nu} - \eta_{\mu \nu}, \\
\theta_{\mu \nu} & \equiv & h_{\mu \nu} - \frac 1 2 \eta_{\mu \nu} h,
\hspace{.5 in} (h=\eta^{\alpha \beta} h_{\alpha \beta}), 
\end{eqnarray}
where $g_{\mu \nu}$ is the metric.  Indeces of all non-tensors are 
raised and lowered with the Minkowski metric 
$\eta^{\alpha \beta}=\eta_{\alpha \beta}={\rm diag} (-1, +1, +1, +1).$  Also,
we define $,_{\nu} \equiv \partial / \partial x^{\nu}$, and take $G=c=1$.  
This allows us to write the Einstein field equations as
\begin{equation}
\eta^{\alpha \beta} \theta^{\mu \nu},_{\alpha \beta}=-16 \pi \tau^{\mu \nu}
\label{Einstein}
\end{equation}
with the gauge condition
\begin{equation}
\theta^{\mu \nu},_{\nu}=0.
\end{equation}
Here we followed Epstein in  defining 
\begin{equation}
\tau^{\mu \nu}=\tau^{\nu \mu}=T^{\mu \nu}+t^{\mu \nu}.
\label{stress-energy}
\end{equation}
$T^{\mu \nu}$ is the matter stress-energy tensor and $t^{\mu \nu}$ is 
the suitably defined gravitational stress-energy pseudo-tensor.

Equation (\ref{Einstein}) is in the form of a wave equation in flat space. 
Its solution for outgoing wave boundary conditions is
\begin{equation}
\theta^{\mu \nu}(t,{\bf x})=4\int \tau^{\mu \nu}(t-|{\bf x}-{\bf x'}|,{\bf x'})
|{\bf x}-{\bf x'}|^{-1} d^3 x'
\label{intsol}
\end{equation}
at any field point $x^{\mu}=(t,{\bf x})=(x^0,x^i),i,j,...=1,2,3$. We will 
assume all gravitational fields are weak enough that we may evaluate equations
(\ref{stress-energy}) and (\ref{intsol}) to leading order in $\theta^{\mu
  \nu}$.

Following Epstein~\cite{Epstein}, let us 
consider a source of the form
\begin{equation}
\tau^{ij}(t, {\bf x})=n^i n^j r^{-2} \sigma(t-r)f(\Omega, t-r), 
\end{equation}
where ${\bf n}={\bf x}/r$, $r=|{\bf x}|$, $f(\Omega,t) \geq 0$, $\int
f(\Omega,t) d\Omega \equiv 1$.  This form describes neutrinos continuously
released from ${\bf x}=0$ at the speed of light.  $\sigma(t)$ and
$f(\Omega,t)$ are the rate of energy loss (luminosity) and angular
distribution, respectively.  In the weak field limit, it is safe to treat
all neutrinos as orriginating from ${\bf x}=0$ and to use $t-r=$const as an
approximation to a null geodesic.

To obtain the detectable radiation we must find the transverse 
traceless (TT) part of the source.  This can be done either with a projector
or with a gauge transformation.  Let us use the former.  
The detectable radiation in then
\begin{equation}
\theta^{\mu \nu}_{TT}(t,{\bf x}) =  
            4\int \tau^{\mu \nu}_{TT}(t-|{\bf x}-{\bf x'}|,{\bf x'})
|{\bf x}-{\bf x'}|^{-1} d^3 x', 
\label{intsolTT}
\end{equation}
where 
\begin{equation}
\tau^{ij}_{TT}=(P^i_k P^j_l - \frac 1 2 P^{ij}P_{kl})\tau^{kl}
\end{equation}
with
\begin{equation}
P^i_j=\delta^i_j - \tilde{n}^i \tilde{n}_j, \hspace{.5 in}
{\bf \tilde{n}}=\frac {\bf x-x'} {|{\bf x-x'}|}.
\end{equation}
Here 
${\bf n}$ is a unit vector in the ${\bf x}$ direction, ${\bf n'}$ is in the
${\bf x'}$ direction and ${\bf \tilde{n}}$ is in the ${\bf x-x'}$ direction.
Let $\theta$ be the angle between $\bf n$ and $\bf n'$, $\psi$ is 
the angle between $\bf n$ and $\bf \tilde{n}$ and $\psi+\theta$ is the angle
between $\bf n'$ and $\bf \tilde{n}$.  

Then assuming ${\bf x}$ is along the $z$-direction, and ${\bf n'}$ has polar 
coordinates $(\theta,\phi)$, we find
\begin{eqnarray}
(n'^x n'^x)_{TT}&=&\frac 1 2 \sin^2(\psi+\theta)
        \left[ \cos^2 \phi (1+cos^2\psi)-1 \right], \\
(n'^y n'^y)_{TT}&=&\frac 1 2 \sin^2(\psi+\theta)
        \left[ \sin^2 \phi (1+cos^2\psi)-1 \right], \\
(n'^x n'^y)_{TT}&=&\frac 
        1 2 \sin^2(\psi+\theta) \sin \phi \cos \phi (1+\cos^2 \psi), \\
(n'^z n'^z)_{TT}&=&\frac 1 2 \sin^2 \psi \sin^2(\psi + \theta), \\
(n'^z n'^x)_{TT}&=&\frac 1 2 \sin \psi \cos \psi \cos \phi 
            \sin^2(\psi+\theta),\\
(n'^z n'^y)_{TT}&=&\frac 1 2 \sin \psi \cos \psi \sin \phi 
            \sin^2(\psi+\theta).\\
\end{eqnarray}
Using the identities 
\begin{equation}
\frac {\sin \psi} {r'}=\frac {\sin \theta} {\bf |x-x'|}=
\frac {\sin (\psi+\theta)} {r}, 
\label{lawsin}
\end{equation}
\begin{equation}
{\bf |x-x'|}^2=r^2+r'^2-2 r r' \cos{\theta}, 
\label{lawcos}
\end{equation}
we write equation (\ref{stress-energy}) as
\begin{equation}
\tau^{ij}(t,{\bf x})=n^i n^j r^{-2} \int_{-\infty}^{\infty}
f(\Omega',t')\sigma(t')\delta(t-t'-r) dt'.
\end{equation}
Inserting this into equation (\ref{intsolTT}) we obtain 
\begin{equation}
\theta^{ij}_{TT}(t,{\bf x})= 4 \int_{-\infty}^{\infty} \int_{4 \pi}
\int_0^{\infty} \frac {(n^i n^j)_{TT} f(\Omega',t') \sigma(t')} {\bf| x-x'|}
\delta(t-{\bf |x-x'|}-t'-r')dr' d\Omega' dt'.
\end{equation}
We can now perform the $r'$ integration using equations (\ref{lawsin}) and 
(\ref{lawcos}) to yield
\begin{equation}
\theta^{ij}_{TT}(t,{\bf x})= 4 \int_{-\infty}^{t-r} \int_{4 \pi}
\frac {(n^i n^j)_{TT} f(\Omega',t') \sigma(t')} {t-t'-r\cos\theta'}
d\Omega' dt'.
\label{intsolTT3}
\end{equation}
Let us define
\begin{equation}
\alpha=\frac {t-t'} {r}-1.
\end{equation}
Then we can write all the angles in terms of $\theta'$ and $\alpha$
$$\sin \psi = \frac {r'} {\bf |x-x'|} \sin \theta'=
\frac {\alpha^2+2\alpha} {\alpha^2+2\alpha(1-\cos\theta')+2(1-\cos\theta')}
\sin{\theta'},$$
$$\sin(\psi+\theta')= \frac {r} {\bf |x-x'|} \sin \theta'=
   \frac {2(\alpha+1-\cos\theta')}
      {\alpha^2+2\alpha(1-\cos\theta')+2(1-\cos\theta')}\sin{\theta'}.$$ 
We can then write 
\begin{equation}
Q^{ij}(\alpha,t')=\int_{4 \pi}
\frac {(n'^i n'^j)_{TT}(\alpha,\Omega') f(\Omega',t')}
      {\alpha+1-\cos\theta'} d\Omega'. 
\end{equation}
Next we write equation (\ref{intsolTT3}) as
\begin{equation} 
\theta^{ij}_{TT}(t,{\bf x})= 4
r^{-1}\int_{-\infty}^{t-r}Q^{ij}(\alpha,t') 
\sigma(t') dt'.
\end{equation}
Let us note that all of the dependence on $t$ (not $t'$) is either in the
integration limit, or in $\alpha$. Thus, we can write the time rate of
change of $\theta^{ij}$ as
\begin{equation}
\theta^{ij}_{,0TT}(t,{\bf x})= 4 r^{-1}Q^{ij}(0,t-r)\sigma(t-r)
+4 r^{-2}\int_{-\infty}^{t-r}\frac {\partial Q^{ij}(\alpha,t')} 
{\partial \alpha} \sigma(t') dt'.
\end{equation}

We see that there are two contributions.  The first is due to the new
neutrinos being ejected at $t-r$, the ones whose effect is just becoming
visible.  The other, is due to the change in the effect of previously
ejected neutrinos, as their distance and angle change.  However, this
second term is suppressed by an extra factor of $r$, so it is negligible as
long as the radiation lasts for a time that is short compared to the
distance between the source and the observer.  (In a supernova, the event
is about 10 s and the distance is thousands of light years so we can safely
ignore this term.)

For the first term, since $\alpha=0$ and $\psi=0$, the angular terms simplify
greatly.  All of the $\theta^{zi}_{TT}$ terms drop out and we find
\begin{eqnarray}
\nonumber (\theta^{xx},_0)_{TT}({\bf x},t)&=& \\ 
-(\theta^{yy},_0)_{TT}({\bf x},t)&=& 2
r^{-1}\sigma(t-r) \int_{4\pi} (1+\cos{\theta'})\cos{2\phi'}f(\Omega',t-r)
d{\Omega'} \label{thetaxx},\\ 
(\theta^{xy},_0)_{TT}&=& 2 r^{-1}\sigma(t-r)
\int_{4\pi} (1+\cos{\theta'})\sin{2\phi'}f(\Omega',t-r) d{\Omega'}.
\label{thetaxy} 
\end{eqnarray}

\section{$\delta$-function Distribution}

At this point we simply need to choose a distribution $f(\Omega',t)$ to
find the form of the radiation.  The simplest is just a $\delta$ function
in a particular direction.  In essence, we are modeling the slight neutrino
excess to one side as a $\delta$-function jet.  In reality the $\delta$
function introduces some spurious results, but it also provides a basis for
more accurate results.  Let
\begin{equation}
f(\Omega',t)=\delta(\cos\theta(t)-\cos\theta'(t))\delta(\phi(t)-\phi').
\end{equation}
Then we find
\begin{eqnarray}
\nonumber \theta^{xx},_{0TT}({\bf x},t)&=& \\
-\theta^{yy},_{0TT}({\bf x},t)&=& 2 r^{-1}\sigma(t-r)
(1+\cos{\theta(t-r)})\cos{2\phi(t-r)},\\
\theta,^{xy}_{0TT}&=& 2 r^{-1}\sigma(t-r)
(1+\cos{\theta(t-r)})\sin{2\phi(t-r)}. 
\end{eqnarray}
The most striking feature is that while the radiation is suppressed for
$\theta=\pi$ when the neutrino jet is away from the the observer as
expected, there is no suppression near $\theta=0$ when the jet is toward
the observer.  We will see later that this is one of the spurious effects
of the $\delta$-function.

The neutrino jet's motion could be quite complex.  We will focus on one
simplified possibility where the jet precesses at an angular frequency
$\omega$ around an axis that is inclined at an angle $\alpha$ relative to
the $z$-axis, and always makes an angle of $\chi$ relative to that axis.
For simplicity we will always assume that the axis of precession is in the
$x-z$-plane.  Rotating it away from this plane can be handled most easily
by rotating the observer's axes.

If $\alpha=0$ then the radiation is simply
\begin{eqnarray}
\theta^{xx},_{0TT}({\bf x},t)&=& 2 r^{-1}\sigma(t-r)
(1+\cos{\chi})\cos{2\omega(t-r)},\\
\theta^{xy},_{0TT}({\bf x},t)&=& 2 r^{-1}\sigma(t-r)
(1+\cos{\chi})\sin{2\omega(t-r)}. 
\end{eqnarray}
Which is just circularly polarized gravitational radiation at an angular
frequency of  
$2\omega$ which is exactly what we would expect.

If we move $\alpha$ away from the $z$-axis.  Things get more complicated.
By simply rotating the axes we can find
\begin{eqnarray}
\cos\theta&=&\cos\alpha \cos\chi-\sin\alpha \sin\chi \cos\phi_0, \\
\tan\phi&=&\frac {\sin\chi \sin\phi_0} {\sin\alpha  \cos\chi + \cos\alpha
  \sin\chi  
\cos\phi_0}, \label{GWequation1}\\
\cos {2 \phi}&=&\frac {1-\tan^2 \phi} {1+\tan^2 \phi}, \\
\sin {2 \phi}&=&\frac {2 \tan \phi} {1+\tan^2 \phi},
\label{GWequation2}
\end{eqnarray}
where we have let $\phi_0=\omega(t-r)$.  Using these gives us gravitational
radiation that is elliptically polarized and and contains components at all
angular frequencies $n \omega$ where $n$ is an integer.  The radiation is
generally dominated by $\omega$ and $2 \omega$ components, and in fact we
will find that the higher frequency terms are spurious effects of the
$\delta$-function.

The magnitude of any particular Fourier component is then
\begin{equation}
\theta^{ij}_{TT} \approx \frac 2 r \frac {e L} {\omega} \sqrt{\tau}
\end{equation}
$L$ is the luminosity of neutrinos, $e$ is the excess in one direction
(about 1\%), $\omega$ is the frequency of the radiation (not the frequency
of precession this time), and $\tau$ is the length of time over which the
neutrinos are released.  (For burst radiation this would be a factor of
$\sqrt{\frac 1 {2 \pi \omega}}$).

\section{Simple Dipole}
The $\delta$-function is of course an idealizaton.  The excess of neutrino
energy in one direction is caused by the magnetic field and is expected to
be of the form
\begin{equation}
f(\Omega',t)=\frac 1 {4 \pi} \left(1+3e\cos{\tilde{\theta}}\right).
\label{alexform}
\end{equation}
This distribution is chosen so that it has only a uniform and a dipole 
contribution.  The numerical factors are chosen so that $\int f d\Omega' =1$
and $\int f \cos{\tilde{\theta}} d \Omega'=e$.  Here $e$ is again the 
fractional excess of neutrino momentum in one
direction compared to the other, and $\tilde{\theta}$ is the angle to the
magnetic axis.

If the magnetic axis points in the $(\theta, \phi)$ direction, and we are
looking in the  $(\theta', \phi')$ direction, then it is easy enough to find 
\begin{equation}
\cos{\tilde{\theta}}=\hat{n}(\theta,\phi) \cdot \hat{n}(\theta',\phi')
=\sin{\theta}\sin{\theta'}(\cos\phi \cos\phi' + \sin\phi \sin\phi') + 
\cos{\theta}\cos{\theta'}.
\label{costildetheta}
\end{equation}

If we now insert equations (\ref{alexform}) and (\ref{costildetheta}) into
equations (\ref{thetaxx}) and (\ref{thetaxy}) each term of the integral over 
$\phi'$ will take one of the following three forms:
\begin{equation}
\int_0^{2\pi}\cos 2\phi' d \phi' =
\int_0^{2\pi}\cos 2\phi' \sin \phi' d \phi'=
\int_0^{2\pi}\cos 2\phi' \cos \phi' d \phi' = 0.
\end{equation}
All three are zero, so the result for $\theta^{xx},_{0TT}$ and 
$\theta^{xy},_{0TT}$ is 
identically $0$.
This is a reflection of the fact that there is no gravitational dipole
radiation.

The first term that will produce noticable radiation is the quadrupole term
which we will write in the form
\begin{equation}
f(\Omega',t) = \frac {\gamma} {4 \pi} \cos^2\tilde{\theta}.
\label{quadform}
\end{equation}
This is actually not a true quadrupole as it has a monopole contribution, 
but that is irrelevant as monopoles will not create GWs.  The factor of $4\pi$
is to remain consistent with equation (\ref{alexform}).  The factor of $\gamma$
represents the strength of this term.  It is left general, but
we would expect that if the dipole term is suppressed by the factor $e$, the
quadrupole term should be suppressed by two factors of e so that 
$\gamma \sim e^2$.
 
If equation (\ref{quadform}) is inserted into 
(\ref{thetaxx}) and (\ref{thetaxy}) then we get
\begin{eqnarray}
\theta^{xx},_{0TT}({\bf x},t)&=& 2 r^{-1}\sigma(t-r)
\frac {\gamma} 6 \sin^2{\theta} \cos{2\phi}, \\
\theta^{xy},_{0TT}({\bf x},t)&=& 2 r^{-1}\sigma(t-r)
\frac {\gamma} 6 \sin^2{\theta} \sin{2\phi}.
\end{eqnarray}
This is nearly the same result as the $\delta$-function case except for the
numerical factor and a $\sin^2\theta$ instead of $1+\cos\theta$.  This
small change results in suppressed radiation whenever the predominiant
neutrino direction is alligned with the observer's line of sight.  (Because
it is a quadrupole term, there is no distinction between aligned and
anti-aligned.)  Even more interesting, if we allow the axis of rotation to
be different from the $z$-axis as we did in the case of the
$\delta$-function, we find that we only have constant terms and terms with
angular frequency $\omega$ and $2\omega$.  All of the higher frequency
terms are absent.  Unfortunately, it seems at first glance that
$\gamma \sim e^2$ so this radiation is likely to be suppressed by two more
orders of magnitude and may be unobservable.

\section{Realistic quadrupole from an off-centered distribution} 

The best chance of observing such gravitational waves is from an object
with a large quadrupole moment.  The pulsar kick mechanism based on neutrino
conversions~\cite{OscKicks,sterile,sterile_dark} predicts an asymmetric
neutrino emission whose strength and direction are determined by the
magnetic field inside the neutron star.  The magnetic fields deep inside
the neutron stars are believed to grow during the first seconds after the
supernova collapse ($\alpha-\Omega$ dynamo effect~\cite{zeldovich}) because
of convection.  The cooling of the outer regions of the star creates the
temperature and entropy gradient that causes convection in these outer
regions.  It is likely, therefore, that the strongest magnetic field
develops away from the central region and that the overall magnetic field has a
large non-dipole component.  Since the neutrino emission is affected by
this magnetic field, the asymmetric part of the neutrinos can form an
off-centered beam.  This, in turn creates a source for GW with a large
quadrupole moment.


Here we consider two cases, which correspond to resonant and off-resonant
conversions of neutrinos in Refs.~\cite{sterile} and~\cite{sterile_dark},
respectively. 

\subsection{Resonant conversions} 

Let us consider a $B$-field of the form 
\begin{equation}
\bf{B}=\hat{k} B_0 {\left( \frac {\rho} {\rho_0} \right)}^2.
\label{Bfield}
\end{equation}
Here $\rho$ is the distance from some magnetic axis, $B_0$ is the magnetic
field at the center of the star and $\rho_0$ is the distance from the
magnetic axis to the center of the star.  This is the simplest model that
has no singularity in the current at the magnetic axis and represents
charged particles circling with a constant angular velocity so that the
current increases linearly with $\rho$. This model is based on cylindrical
symmetry so it is a simplification for our case, but it should be
sufficient for an order of magnitude estimate.

According to Ref.~\cite{OscKicks,sterile}, the radius at which neutrinos
escape can be described as
\begin{equation}
r=r_0+\delta r,
\end{equation}
\begin{equation}
2 \frac {\partial N_e} {\partial r} \delta r = -e {\left( \frac {3 N_e} 
{\pi^4}\right)}^{1/3} \frac {\bf k \cdot B} k.
\end{equation}
The only change from the reference is that I'll allow $B$ to vary with
distance from the magnetic axis.  For simplicity I'll place the magnetic
axis in the $x-z$ plane.  Then we can find $\delta$ as a function of
$\tilde{\theta}$ and $\tilde{\phi}$:
\begin{equation}
\delta r = \delta_0 \cos \tilde{\theta} 
[1 - 2x \sin \tilde{\theta} \cos\tilde{\phi}+x^2\sin^2\tilde{\theta}];
\label{realquad2}
\end{equation}
$\delta_0$ is the same as $\delta$ in the reference (except with the
assumed constant B-field replaced by the B field at the origin.)  The value
$x$ is simply the ratio of the unperturbed escape radius to the distance
between the magnetic axis and the center of the star $x=\frac
{r_0} {\rho_0}$.  Both distances can be a significant fraction of the star's
overall size and are not correlated so $x$ is most likely of order 1 and can 
be greater or less than 1.

The first term is the one discussed in Ref.\cite{OscKicks,sterile} it
is responsible for the momentum kick and as demonstrated causes no
gravitational radiation.  The last term is mainly $\ell=3$ and also
doesn't contribute to the gravity waves.  The middle term however, is a
quadrupole term and does create gravity waves.

Its worth noting that while this is actually a measure of how much the
neutrinosphere deviates from the sphere, it is assumed that the neutrinos,
responding to their environment, will have an energy spectrum that mimics
the radius distribution.
%
%
\begin{equation}
f(\Omega',t)=
      \frac 1 {4 \pi} \left(1+3e(1+\frac 2 5 x^2)^{-1}\cos \tilde{\theta} 
[1 - 2x \sin \tilde{\theta} \cos\tilde{\phi}+x^2\sin^2\tilde{\theta}]\right). 
\label{realquad}
\end{equation}
This distribution has been chosen to have a large monopole term, and a 
perturbing term of the same form as equation (\ref{realquad2}).  Like 
equation (\ref{alexform}) it has been normalized so that
$\int f d\Omega' = 1 $ and $\int f \cos {\tilde{\theta}} d\Omega'=e$ 
where $e$ represents the fractional momentum excess
in one direction.

The next task is to find $\cos \tilde{\theta} \sin{\tilde{\theta}}
\cos{\tilde{\phi}}$ in terms of $(\theta, \theta', \phi, \phi')$.  As I
already discussed, $\cos{\tilde{\theta}}$ is the projection of the
$\hat{r}'$ ($\hat{n}'$) vector (the integration direction) onto the $\hat{r}$ 
($\hat{n}$)vector
(the direction of the magnetic field axis).  Essentially, the radial
direction is the z direction for the tilde variables, and
$\cos{\tilde{\theta}}$ is the projection onto this axis.

By contrast $\sin{\tilde{\theta}}\cos{\tilde{\phi}}$ is the projection onto
the $x$-axis for the tilde variables.  Since $\hat{r}$ is the
$\tilde{z}$-direction, the $\tilde{x}$-direction has to be in the
$\hat{\theta}$, $\hat{\phi}$ plane, but it can be any linear combination of
these.  So I'll let it be
\begin{equation}
\tilde{\hat{x}}=\cos{\alpha}\hat{\theta}+\sin{\alpha}\hat{\phi}.
\end{equation}  
In terms of the standard $(x,y,z)$ directions we can write
\begin{eqnarray}
\hat{\theta}&=&\left( \begin{array} {c} \cos{\theta}\cos{\phi} \\
  \cos{\theta}\sin{\phi} \\  
-\sin{\theta}
\end{array} \right), \\
\hat{\phi}&=&\left( \begin{array} {c} -\sin{\phi} \\ \cos{\phi} \\ 0
\end{array} \right), \\
\hat{r'}&=&\left( \begin{array} {c} \sin{\theta'}\cos{\phi'} \\
  \sin{\theta'}\sin{\phi'} \\  
\cos{\theta'}
\end{array} \right).
\end{eqnarray}
Which means
\begin{equation}
\sin{\tilde{\theta}}\cos{\tilde{\phi}}=\cos{\alpha}\left( 
\cos{\theta}\sin{\theta'}\cos{(\phi-\phi')}-\sin\theta \cos{\theta'}\right)
+ \sin{\alpha}  
\left(\sin{\theta'} \sin{(\phi'-\phi)}\right),
\label{sintilde}
\end{equation}
\begin{equation}
\cos{\tilde{\theta}}=\sin{\theta}\sin{\theta'}\cos{(\phi-\phi')}-
\cos{\theta}\cos{\theta'}.  
\label{costilde}
\end{equation}
Substituting this into equation \ref{realquad} and in turn into 
the GW equations (\ref{thetaxx} and \ref{thetaxy}) 
we get
\begin{eqnarray}
\left(\begin{array} {c} \theta^{xx},_{0TT}({\bf x},t)\\
\theta^{xy},_{0TT}({\bf x},t)\end{array}\right)& = & 
-\frac{2 e \sigma(t-r)}{ 
r \left (1+\frac 2 5 x^2 \right)}
\sin{\theta}  
\left( \begin{array} {c} 
\cos\theta \cos\alpha \cos{2\phi} - \sin{\alpha} \sin{2\phi} \\
\cos\theta \cos\alpha \sin{2\phi} + \sin{\alpha} \cos{2\phi} \end{array}
\right) \nonumber \\ 
& = & -\frac{2 e \sigma(t-r)}{ 
r \left (1+\frac 2 5 x^2 \right)}
\sin{\theta} 
\sqrt{\cos^2 \theta \cos^2 \alpha+\sin^2\alpha} 
\left( \begin{array} {c} \cos{(2\phi+\delta)} \\ \sin{(2 \phi+\delta)}
\end{array}  \right).    
\label{RealQuadGW} 
\end{eqnarray}
Here $\tan{\delta}=\tan{\alpha}\sec{\theta}$. 

Clearly there is radiation and there is no extra suppression compared to
the dipole term.  There is now a wider variety of radiation producing
motions.  Not only can we allow $\theta$ and $\phi$ to change based on
rotation about some axis, but $\alpha$ can also change.  However, if we
assume a simple rotation of angular frequency $\omega$ then $\alpha$ will
be linked to the rotation angle.  There can be a phase difference between
them $\alpha=\phi_0+\epsilon$, but $\epsilon$ goes to 0 if the
magnetic and rotation axes are coplanar or $\pi/2$ when maximally skewed.
There is now an extra factor of the rotation angle which adds significant
radiation at a frequency of $3\omega$ as well as $\omega$ and $2\omega$.

The maximally skewed case is to be expected if the neutrino kick is also
driving the neutron star's rotation as in \cite{RotKicks}.  This could also
suppress the translational kick and make the gravity waves larger than
expected relative to the proper motion.

\subsection{Off-resonant neutrino conversions}

If the active-to-sterile neutrino conversions occur off-resonance, the
sterile neutrinos are emitted from the entire core of the
star~\cite{sterile_dark}.  Such emission is only possible after the
electroweak matter potential is driven to zero by neutrino
oscillations~\cite{Fuller}.  Therefore, in this case, the sterile emission
can begin several seconds after the on  set of the supernova.  More
importantly, this emission may last for a period of time that could be as
long as 10 s, but can also be much shorter.  This depends on the neutrino
mass and mixing parameters.  

The emission of neutrinos in the solid angle $d \Omega$ is $\frac {dN}
{d\Omega}=N_0(1+\epsilon\cos\tilde{\theta})$. 
The dependence on the magnetic field is more complicated than in previous
models.
However, we can get a reasonable order of magnitude estimate by parametrizing 
the emission in the same way as (\ref{realquad2}). 
Specifically the gravity wave generating term will take the form
\begin{equation}
-2 N_0 \epsilon_0 x \sin{\tilde{\theta}}\cos{\tilde{\theta}}\cos{\tilde{\phi}}.
\end{equation}
This must be integrated over all radii where neutrinos are emitted.
Assuming they are uniformally emitted throughout the core (i.e. no explicit 
dependence on radius) the final form will be 
\begin{equation}
-\frac 3 2 N_0 \epsilon_0 x_{max} \sin{\tilde{\theta}}\cos{\tilde{\theta}}.
\cos{\tilde{\phi}}
\end{equation}

Except for the extra factor of $3/4$ the result is the same as for the resonant
case and will yield a similar magnitude signal.

\section{Signal Magnitude}
\label{sigmag}
We are now ready to consider the magnitude of the radiation.  In a typical
supernova, neutrinos carry away about $10^{53}$ ergs over a period of several 
seconds.  So $\sigma$ is approximately the average luminosity 
$\sigma=L=10^{52} (10 {\rm s}/\tau)$ erg/s where $\tau$ is the total time of
neutrino emmision.  The magnitude of the gravity waves is then
\begin{equation}
\theta,_{0TT}(t)\approx \left( \frac {2 e \times 10^{52} {\rm erg/s}} {r} 
                    \right)
                    \left( \frac {10 {\rm s}} {\tau} \right) F(\theta_i,t).
\label{mag1}
\end{equation}
$F(\theta_i,t)$ is the function of angles appearing in equation 
(\ref{RealQuadGW}).  $\theta_i$ merely represents the collection of angles.
 
To be compared with experiment, this result needs to be integrated
with respect to time and Fourier transformed.  In the Fourier transform all 
three major components ($\omega$, $2\omega$, and $3\omega$) have amplitudes 
of order $1$, and integrating is equivalent to dividing by the angular 
frequency.  

The units must be changed also.  The 
gaussian noise must be Fourier transformed in a way that gives it units of
${\rm Hz}^{-1/2}$.  To adjust for this we must divide by the square root of 
the frequency for a burst source, or multiply by the square root of $\tau$ for
a periodic source  \cite {Thorne}.
Since neutron stars have typical frequencies of several 
kHz, and the burst is expected to last a few seconds, we should generally treat
this as a periodic source.  However, in the case of lower frequency signals 
detectable by LISA, the burst description may be more appropriate.

The final result is 
\begin{equation}
\theta_{TT}(f) \approx 10^{-24} {\rm Hz}^{-1/2}
\left(\frac {e} {.01} \right) {\left( \frac {10 {\rm s}} {\tau} \right)}^{1/2}
\left(\frac {\rm 1 kpc} r \right) \left( \frac {\rm 1 kHz} f \right)
\label{periodic}
\end{equation}
for periodic signals, or
\begin{equation}
\theta_{TT}(f) \approx 3 \times 10^{-22} {\rm Hz}^{-1/2}
\left(\frac {e} {.01} \right) {\left( \frac {10 {\rm s}} {\tau} \right)}
\left(\frac {\rm 1 kpc} r \right) {\left( \frac {\rm 1 Hz} f \right)}^{3/2}
\label{burst}
\end{equation}
for a burst signal.
In both cases $f$ represents the frequency of the radiation, not the star 
rotation.  (Thus, $f$ will be 1, 2, or 3 times the frequency of star rotation.)

\begin{figure}
\begin{center}
\includegraphics[height=7cm]{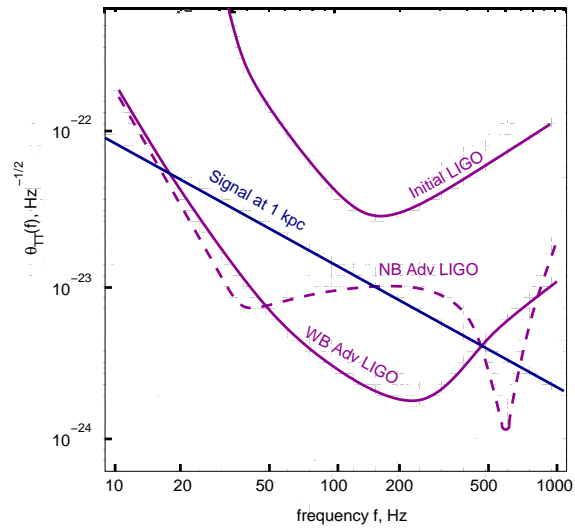}%
\end{center}
\caption{Gravity wave intensity from a pulsar kick caused by
neutrino oscilations.  The LIGO sensitivity is shown for reference.  
For near supernovas with the appropriate rotation frequency, the signal should
be observable.  Please note this shows the magnitude at a given frequency.  
The actual signal is not continuous, but would be concentrated at three 
closely spaced frequencies, $f$, $2f$, and $3f$.}
\label{ligofig}
\end{figure}

\begin{figure}
\begin{center}
\includegraphics[height=7cm]{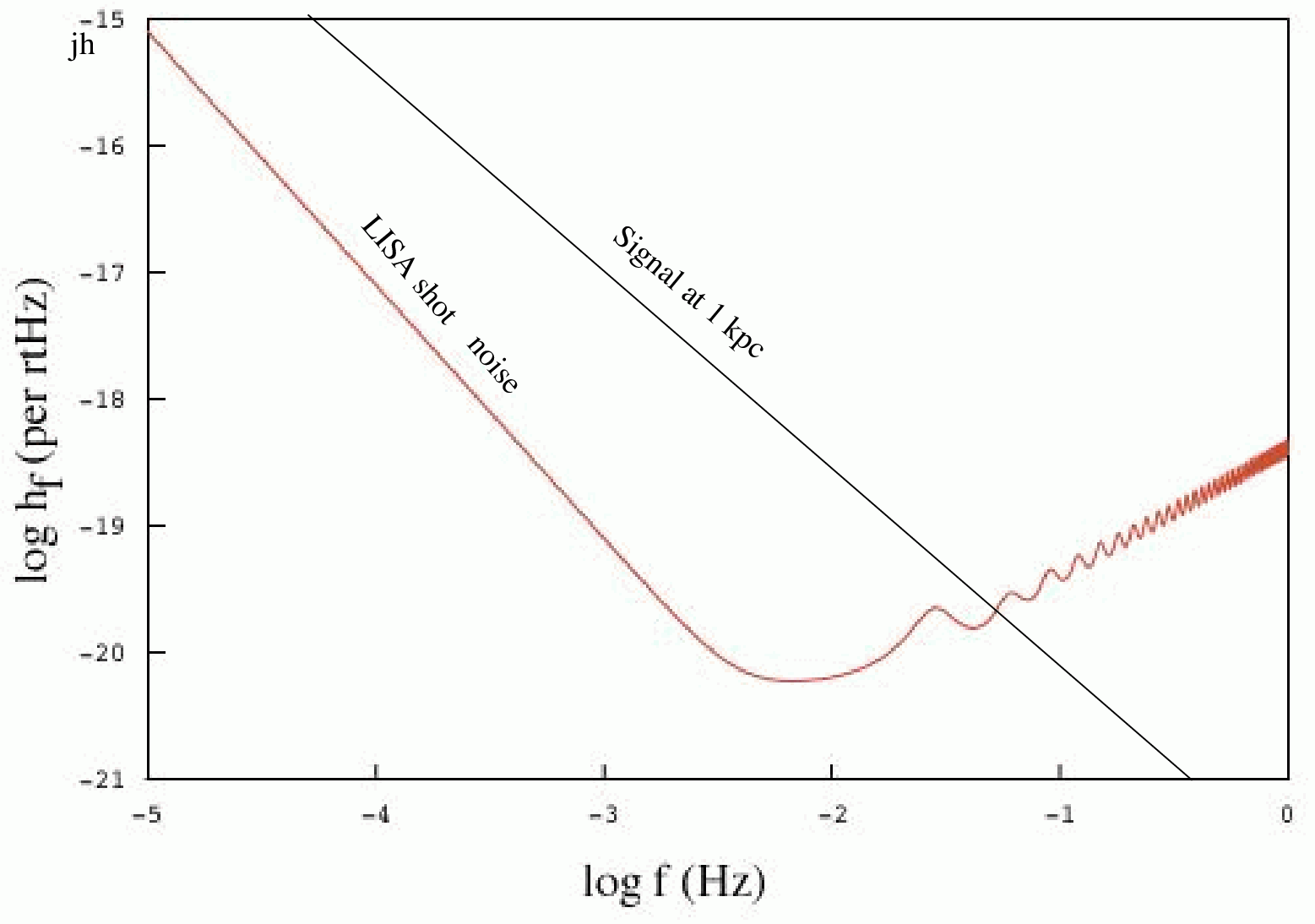}%
\end{center} 
\caption{Gravity wave intensity from a pulsar kick caused by
neutrino oscilations.  The LISA sensitivity is shown for reference.  Note that
for near supernovas with the appropriate rotation frequency, the signal can 
be observable. Please note this shows the magnitude at a given frequency.  
The actual signal is not continuous, but would be concentrated at three 
closely spaced frequencies, $f$, $2f$, and $3f$.}
\label{lisafig}
\end{figure}

LIGO is most sensitive in the $10$ - $1000$ Hz range, neutron stars 
rotating at this rate are likely to undergo several rotations in the 
neutrino ejection time and therefore equation (\ref{periodic}) is most 
appropriate in describing LIGO's sensitivity to gravity waves figure 
\ref{ligofig}. 
 LISA by contrast is most sensitive in the $.001$ - $.1$ Hz range.  Neutron 
stars rotating at this rate are likely to undergo only part of a 
rotation, or at most a few rotations in the neutrino ejection time.  
Thus, equation (\ref{burst}) is more appropriate in describing its 
sensitivity figure \ref{lisafig}. 

From the figures it appears that LISA is more likely to see this GW signal, 
being sensitive to supernovas as far as several Mpc. Unfortunately, LISA's 
frequency range is very low and it would only see GWs from very slowly 
rotating neutrino beams. 


\section{Gravitational Energy}
The total power and energy carried by the the gravity wave are simply
\begin{eqnarray}
P&=&{|{\theta^{ij}},_{0TT}|}^2, \\
\Delta E&=& {|{\theta^{ij}},_{0TT}|}^2 \tau, 
\end{eqnarray}
where $\theta^{ij},_{0TT}$ can be given by any of the several equations for
this quantity as appropriate.  Inserting equation (\ref{mag1}) this means
that the energy carried away by gravity waves will be about $.1$ erg /
cm$^2$ or a total of about $10^{43}$ erg.

By comparison, M\"uller and Janka \cite{Janka} find a GW energy of
$E^{\nu}_{\rm GW} \sim [10^{-10} - 10^{-13}]M_{\odot}c^2 \sim [10^{44} -
10^{41}] {\rm erg}$, for a luminosity of $L_{\nu} \sim 10^{53} {\rm erg/s}$.
Cuesta \cite{Cuesta2} gets a value of 
$E^{\nu}_{\rm GW} \sim 10^{-4} M_{\odot}c^2 \sim [10^{50}] {\rm erg}$ with 
a luminosity of $L_{\nu} \sim 10^{56} {\rm erg/s}$.  (The neutrino luminosity
is important because gravitational luminosity scales with $L_{\nu}^2$ and the
total energy scales like $L_{\nu}^2 \tau \sim L_{\nu}$.)  Thus, M\"uller and
Janka's estimate is of roughly the same size as mine, Cuesta's is about 3
orders of magnitude higher.  

Furthermore, the GW signal in this paper can be distinguished from other
GWs caused by neutrino emission from a neutron star, by its strong periodic
nature.  This signal will be highly concentrated at three main frequencies
$f$, $2f$, and $3f$ (see section \ref{sigmag}). 
By contrast, other signals have their power spread out
over a large, fairly continuous range of frequencies. In certain ideal 
situations individual cycles of the radiation may be recognizable. This  
should allow the periodic signal to be 
distinguished from other processes including those that contain more total 
energy.

\section{Conclusions}

We have shown that, in the event of a nearby supernova, gravitational waves
caused by an asymmetric neutrino emission may be strong enough to be
detected by LIGO and LISA.  We have assumed that the asymmetry in outgoing
neutrinos is as large as 1\%, as suggested by the observed pulsar
velocities.  As can be seen from the figures, quickly rotating neutron
stars may be seen by LIGO advanced, and more slowly rotating ones by LISA.
This signal also has the distinguishing feature of having large radiation
components at $f$, $2f$, and $3f$.  This could be used for recognizing
gravity waves from rotating neutrino beams.  This mechanism could also be 
recognized by its short time duration and possible coincidence with neutrino 
and optical signals from the supernova.  
 
If such a signal is detected and identified, it would have profound
implications for one's understanding of the supernova, as well as neutrino
physics.  In particular, LIGO and LISA may be able to test the existence
of a keV sterile neutrino, which has been suggested as the explanation
of both the pulsar kicks and the cosmological dark matter.

\section{Acknowledgments}

This work was supported in part by the DOE Grant DE-FG03-91ER40662 and the
NASA grant ATP02-0000-0151.  The author thanks A.~Kusenko, S.~Pascoli, and
H.M.~Cuesta for useful discussions, help and insight.

\end{document}